\documentclass[aps,prl,reprint,amsmath,amssymb,superscriptaddress,showpacs]{revtex4-1}
\usepackage{bm}
\usepackage{graphicx}
\usepackage[usenames, dvipsnames]{color}
\usepackage{dcolumn}
\usepackage{kotex}
\usepackage{float}
\usepackage{hyperref}

\begin{document}

\title{Magnetic excitations in bulk multiferroic two-dimensional triangular lattice antiferromagnet (Lu,Sc)FeO$_3$}

\author{J. C. Leiner}
\email{jleiner@snu.ac.kr}
\affiliation{Center for Correlated Electron Systems, Institute for Basic Science (IBS), Seoul 08826, Republic of Korea}
\affiliation{Department of Physics and Astronomy, Seoul National University, Seoul 08826, Republic of Korea}

\author{Taehun Kim}
\affiliation{Center for Correlated Electron Systems, Institute for Basic Science (IBS), Seoul 08826, Republic of Korea}
\affiliation{Department of Physics and Astronomy, Seoul National University, Seoul 08826, Republic of Korea}

\author{Kisoo Park}
\affiliation{Center for Correlated Electron Systems, Institute for Basic Science (IBS), Seoul 08826, Republic of Korea}
\affiliation{Department of Physics and Astronomy, Seoul National University, Seoul 08826, Republic of Korea}

\author{Joosung Oh}
\affiliation{Center for Correlated Electron Systems, Institute for Basic Science (IBS), Seoul 08826, Republic of Korea}
\affiliation{Department of Physics and Astronomy, Seoul National University, Seoul 08826, Republic of Korea}

\author{T. G. Perring}
\affiliation{London Centre for Nanotechnology and Department of Physics and Astronomy, University College London, 17-19 Gordon Street, WC1H OAH UK}
\affiliation{ISIS Facility, STFC, Rutherford Appleton Laboratory, Didcot, Oxfordshire OX11-0QX, United Kingdom}

\author{H. C. Walker}
\affiliation{ISIS Facility, STFC, Rutherford Appleton Laboratory, Didcot, Oxfordshire OX11-0QX, United Kingdom}

\author{X. Xu}
\author{Y. Wang}
\author{S.-W. Cheong}
\affiliation{Rutgers Center for Emergent Materials and Department of Physics and Astronomy, Rutgers University, Piscataway, New Jersey 08854, USA}

\author{Je-Geun Park}
\email{jgpark10@snu.ac.kr}
\affiliation{Center for Correlated Electron Systems, Institute for Basic Science (IBS), Seoul 08826, Republic of Korea}
\affiliation{Department of Physics and Astronomy, Seoul National University, Seoul 08826, Republic of Korea}

\date{\today}

\begin{abstract}
Non-collinear two-dimensional triangular lattice antiferromagnets (2D TLAF) are currently an area of very active research due to their unique magnetic properties, which lead to non-trivial quantum effects that experimentally manifest themselves in the spin excitation spectra. Recent examples of such insulating 2D TLAF include (Y,Lu)MnO$_3$, LiCrO$_2$, and CuCrO$_2$. Hexagonal LuFeO$_3$ is a recently synthesized 2D TLAF which exhibits properties of an ideal multiferroic material, partially because of the high spin ($S=5/2$) and strong magnetic super-exchange interactions. We report the full range of spin dynamics in a bulk single crystal of (Lu$_{0.6}$Sc$_{0.4}$)FeO$_3$ (Sc doping to stabilize the hexagonal structure) measured via time-of-flight inelastic neutron scattering. Modeling with linear spin wave theory yields a nearest neighbor exchange coupling of $J$~=~4.0(2)~meV (DFT calculations for $h$-LuFeO$_3$ predicted a value of 6.31~meV) and anisotropy values of $K_D$~=~0.17(1)~meV (easy plane) and $K_A$~=~-0.05(1)~meV (local easy axis). It is observed that the magnon bandwidth of the spin wave spectra is twice as large for $h$-(Lu,Sc)FeO$_3$ as it is for $h$-LuMnO$_3$.  
\end{abstract}

\maketitle

\section{Introduction}

The field of multiferroic materials has garnered significant attention in recent years due in part to the promise of potential functionality in advanced information storage and processing applications. This anticipated utility originates from the cross-coupling of electric and magnetic degrees of freedom in phases with simultaneous ferroelectric and magnetic order. On a fundamental level, the coupling between key elementary excitations in the crystal and magnetic lattices, phonons and magnons, plays a key role in driving the phenomena seen in multiferroic materials. Therefore, quantifying the features and effects arising from the correlations between these quasiparticles is of interest from both a practical and fundamental perspective.

Non-collinear two-dimensional triangular lattice antiferromagnets (2D TLAF) have emerged as one possible avenue to the realization of muliferroic materials for practical applications \cite{cheong2007multiferroics, RPP_Tokura}.  Thus, they have attracted the most extensive focus in experimental and theoretical studies. Hexagonal rare-earth manganites (RMnO$_3$) have been a source of much experimental progress in characterizing 2D TLAF. Previous inelastic neutron scattering studies reported that the magnetic excitations of RMnO$_3$ with nonmagnetic R ions (Y/Lu) \cite{Joosung_PRL_13,PhysRevB.68.014432, oh2016} as well as LiCrO$_2$ \cite{toth2016electromagnon} and CuCrO$_2$ \cite{PhysRevB.94.104421} exhibit several anomalous features due to magneto-elastic coupling. In particular, it was explored in detail exactly how spin waves are perturbed by magnon-magnon \cite{PhysRevB.79.144416,RevModPhys.85.219} and magnon-phonon couplings \cite{oh2016, PhysRev.139.A450}. Specifically, a roton-like minimum in the dispersion at the B point \textbf{Q}=($\frac{1}{2}\frac{1}{2}$0) and broadened energy widths at high energy transfers were shown to be present, indicative of the decay of both magnon modes and magneto-elastic hybrid magnon-phonon modes. Careful examination of the data and comparison to theory allowed for quantification of the exchange-striction coupling term. The presence of this magneto-elastic excitation has also been recently corroborated by inelastic X-ray scattering (IXS) measurements \cite{YMOpreparation}.

Such features would naturally be expected to appear in LuFeO$_3$, the Fe based counterpart to LuMnO$_3$. With one more electron in the e$_g$ manifold, it is anticipated that the elaborate effects previously observed in the spin wave spectra of (Y,Lu)MnO$_3$ will be significantly affected by switching the cation at the B site in ABO$_3$ from Mn$^{3+}$ to Fe$^{3+}$ \cite{GaFeO3,das2014bulk}. Effects of this replacement include a significant increase in the magnetic moment and super-exchange interactions, and it is also expected that the magnon-magnon/phonon couplings are rather sensitive these kind of sample parameters \cite{PhysRevB.92.035107}. 

Both forms of LuFeO$_3$, the stable distorted cubic (orthorhombic) $o$-LuFeO$_3$ \cite{Chowdhury_APL,chowdhury2017origin} and the metastable hexagonal $h$-LuFeO$_3$, have previously been studied with the aim of understanding how to harness their properties for logic and memory applications. It has been shown recently that LuFeO$_3$ may be forced into its metastable hexagonal form when it is grown via thin film epitaxy on the appropriate hexagonal crystal substrate \cite{2012epitaxialLFO,bossak2004xrd,Zhang2017}. It is this hexagonal form of LuFeO$_3$ which has been found to exhibit phases of magnetic and ferroelectric order. 

\begin{figure*}
	\centering
	\includegraphics[width=1.0\textwidth,clip]{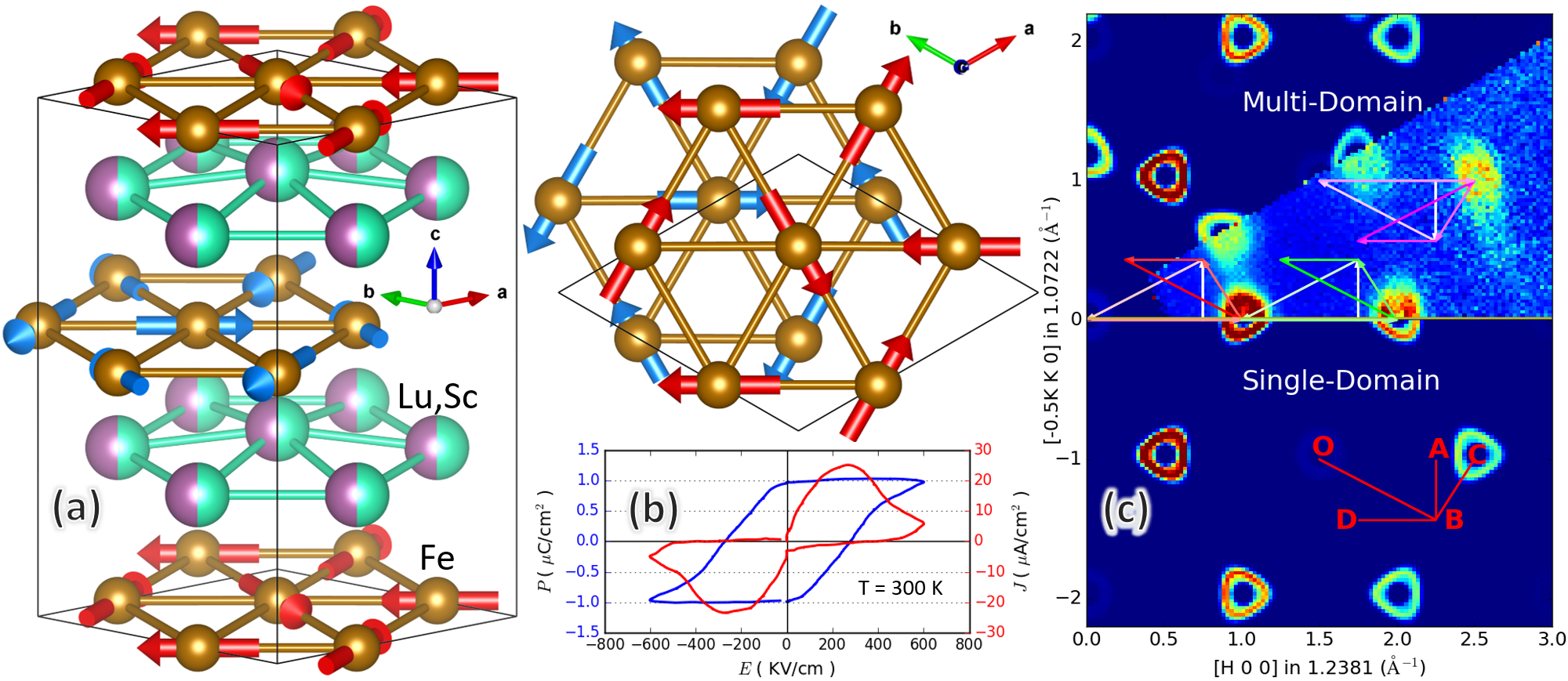}
	\caption{(a) (Lu,Sc)FeO$_3$ crystal (P6$_3$\textit{cm} space group) and magnetic structure. Red (blue) arrows show the configuration in the $z/c$ = 0 ($z/c$ = 1/2) plane of A$_1$ ($\Gamma_1$) the ground state magnetic structure. (b) Hysteresis loops of electric polarization $P$ (blue) and compensated current density $J$ (red) vs. electric field ($E$~$||$~$c$-axis) for the (Lu$_{0.6}$Sc$_{0.4}$)FeO$_3$ crystal used in this study \cite{fn2}. (c) Layout of the in-plane spin waves (from linear spin wave theory) at 10 meV energy transfer, showing the positions of momentum space labels of the high symmetry points (ABCDO) for the triangular lattice. Also shown are the differences in \textbf{Q}-space of single domain and multi-domain \cite{footnote1} cases (domain intensity ratio is $\sim$~1:2). Inset (i.e. the wedge in the lower right of the multi-domain section) shows symmetrized neutron scattering data (integrated in energy transfer from $9~{\rm meV}<\hbar\omega<11~{\rm meV}$) from our (Lu$_{0.6}$Sc$_{0.4}$)FeO$_3$ sample for comparison.  }
	\label{structure}
\end{figure*}

Much of the prior literature on $h$-LuFeO$_3$ is focused on investigating claims which indicate that it is a rare example of a coveted room-temperature multiferroic material \cite{PhysRevB.90.014436,PhysRevLett.110.237601,Chowdhury_APL}, in league with BiFeO$_3$ \cite{Jeong_2012_PRL}. However, a number of studies have reached conclusions that such room-temperature multiferroicity in this material are unfounded, and that in fact the AFM ordering temperature occurs at 155~K, not 443~K \cite{PhysRevB.90.014436,PhysRevLett.114.217602}. Even so, it has recently been shown that in hexagonal Lu$_{1-x}$In$_{x}$FeO$_3$ with $x \approx 0.5$, ferroelectric and antiferromagnetic order are both present at 300~K \cite{RoomTemp_Indium}. Thus, there remains promise that this compound may offer an avenue to a room-temperature multiferroic. 

In order to grow $h$-LuFeO$_3$ in bulk form, it has been shown that instead of using a hexagonal substrate, partial Sc substitution for Lu, Lu$_{1-x}$Sc$_{x}$FeO$_3$, may serve to stabilize the hexagonal structure. This Sc doping increases the AFM ordering temperature as well as the c/a ratio \cite{PhysRevB.92.054435,containerless}. However, multiferroic properties such as the noncollinear magnetic order are not affected \cite{Sc_stabilization}. This is certainly not the case if there is partial Mn substitution on the Fe site: it has been demonstrated that the magnetic structure as well as other properties do indeed change significantly in the case of LuMn$_{0.5}$Fe$_{0.5}$O$_3$ \cite{LMFO}.

Determination of the full spin Hamiltonian for $h$-LuFeO$_3$ is necessary in order to discern the various effects on the spin dynamics outlined above. In this study, we report the full magnetic excitation spectrum of (Lu$_{0.6}$Sc$_{0.4}$)FeO$_3$ in order to understand how the magnetism of the Fe$^{3+}$ ion changes the spin dynamics compared with the hexagonal manganite system LuMnO$_3$. 

\section{Experiment}

So far, a report of the spin dynamics in a single crystal of (Lu,Sc)FeO$_3$ measured via inelastic neutron scattering (INS) is lacking, as such single crystals are rather difficult to grow in bulk form of a sufficient size for INS experiments. However, there have been reports of INS measurements done on a powder sample of (Lu$_{0.5}$Sc$_{0.5}$)FeO$_3$ \cite{PhysRevB.92.054435,Powder_INS}, which determined the magnetic ordering in the ground state to be of A$_1$ ($\Gamma_1$) below $\sim$50~K. The A$_1$ magnetic structure is shown in Fig~\ref{structure}, where the spins are oriented tangentially clockwise around the star pattern seen when the c-axis points out of the page. Above 50~K there is a spin reorientation transition to the A$_2$ ($\Gamma_2$) structure, where the spins are oriented radially outward from the center of the aforementioned star pattern. These and the other different types of magnetic configurations possible for these systems are pictorially presented in several articles~\cite{fiebig2003spin, sim2016hexagonal, das2014bulk}.

Single crystal growth of Lu$_{1-x}$Sc$_{x}$FeO$_3$ has previously been attempted using methods such as containerless processing \cite{containerless}. Recently, a successful synthesis of a $\sim$13 gram single crystal of (Lu$_{0.6}$Sc$_{0.4}$)FeO$_3$ (in the hexagonal P6$_3$\textit{cm} configuration as shown in Fig. \ref{structure}(a)) has been achieved. Doping with at least 40\% Sc has been shown to be necessary to achieve a stabilized hexagonal phase such that a bulk single crystal may be adequately formed.  The (Lu$_{0.6}$Sc$_{0.4}$)FeO$_3$ crystal was grown in an optical floating zone furnace equipped with lasers instead of Halogen lamps. Stoichiometric and high-purity powders of Lu$_2$O$_3$, Sc$_2$O$_3$ and Fe$_2$O$_3$ were mixed in a mortar, pelletized and sintered at 1200$^{\circ}$C for 10 hours. The pellet was ground, re-pelletized and sintered at 1380$^{\circ}$C for 10 hours. The second-sintered pellet was pulverized, poured into a rubber tube and pressed into a rod shape under 8000 PSI hydrostatic pressure. The compressed rod was sintered at 1380$^{\circ}$C for 10 hours. The crystal was grown at the speed of 1 mm/hour in 0.5 MPa O$_2$ atmosphere. The as-grown crystal was annealed at 1400$^{\circ}$C for 20 hours, then cooled down to 1200$^{\circ}$C and room temperature at a rate of 2$^{\circ}$C/hour and 100$^{\circ}$C/hour, respectively. For this sample of $h$-(Lu$_{0.6}$Sc$_{0.4}$)FeO$_3$, the crystallographic parameters are $a$ = $b$ = 5.86 , $c$ = 11.7, and $\alpha$,$\beta$,$\gamma$ = [90$^{\circ}$~90$^{\circ}$~120$^{\circ}$]. As displayed in Fig. \ref{structure}(b), this single crystal has a clear $P(E)$ hysteresis loop at 300~K, unambiguously confirming its ferroelectric nature.

We selected a piece of the floating zone grown sample which was $\sim$ 4~g in mass and had the lowest overall mosaic (estimated to be more than 6$^{\circ}$ on average). It was used for all the measurements described herein. INS measurements via the time-of-flight (TOF) method were collected using the MERLIN spectrometer \cite{MERLIN_instrument,experiment_data_doi} at the ISIS pulsed neutron source in the Rutherford Appleton Laboratory. The sample was placed in a closed-cycle refrigerator (CCR) at the base temperature of 5~K throughout the measurements. 

The optimal balance of intensity and energy resolution for the purposes of this experiment was achieved by setting the Fermi chopper frequency at 350~Hz. A further optimization in the trade-off between energy coverage and resolution was obtained with an incident neutron energy of $E_i$~=~45~meV. The multirep mode \cite{RUSSINA2009624} was utilized to simultaneously obtain additional data at $E_i$ = 24 and 111~meV. The crystal was aligned with the ($HK$0) scattering plane horizontal, meaning that the ($00L$) direction was imaged vertically on the detectors. All of the collected data was reduced, processed, and symmetrized with the Horace software package \cite{Horace_ref}. With the statistics obtained, it is possible to resolve the full spin wave dispersion despite the suboptimal mosaic and 30$^{\circ}$ domain \cite{footnote1} present in this sample (see Fig.~\ref{structure}(c)). The 6$^{\circ}$ mosaic of this sample is the very likely the dominant cause of the unusually broadened spin wave excitations (broadened more than would be expected from the instrument energy resolution of $\sim$2.5~meV FWHM at the elastic line for $E_i$~=~45~meV) seen in the final processed INS data shown in Fig~\ref{Eslices}(a).

\section{Results}

\begin{figure*}
	\centering
	\includegraphics[width=1.0\textwidth,clip]{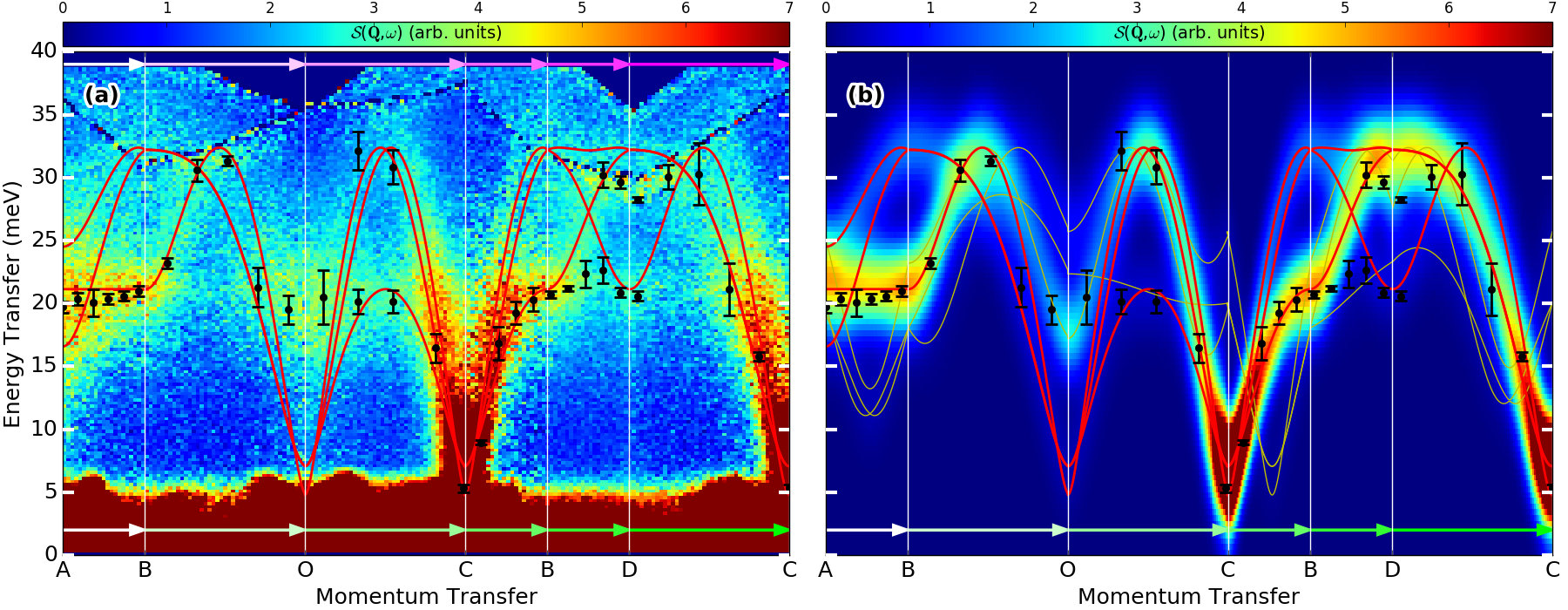}
	\caption{(a) Neutron scattering intensity associated with spin waves (the dynamical spin correlation function $\cal{S}$(\textbf{Q},$\omega$)) for \mbox{$h$-(Lu$_{0.6}$Sc$_{0.4}$)FeO$_3$} experimentally measured at 5~K and (b) calculated from linear spin wave theory (LSWT). $\cal{S}$(\textbf{Q},$\omega$) is plotted along the high symmetry directions as shown in Fig. \ref{structure}(c) ( The dark line in experimental $\cal{S}$(\textbf{Q},$\omega$) data above the highest dispersion branches indicates the boundary between different Brillouin zones as indicated in Fig~\ref{structure}(c). Data below the line corresponds to the paths traced out by the green arrows in Fig~\ref{structure}(c) while data above the line corresponds to purple arrow paths). Red (yellow) lines are LSWT dispersions from the first (second) domain. Black circles are fitting positions from constant-\textbf{Q} cuts through the neutron data ~\cite{SIref}. }
	\label{Eslices}
\end{figure*}

\begin{figure}
	\centering
	\includegraphics[width=1.005\columnwidth,clip]{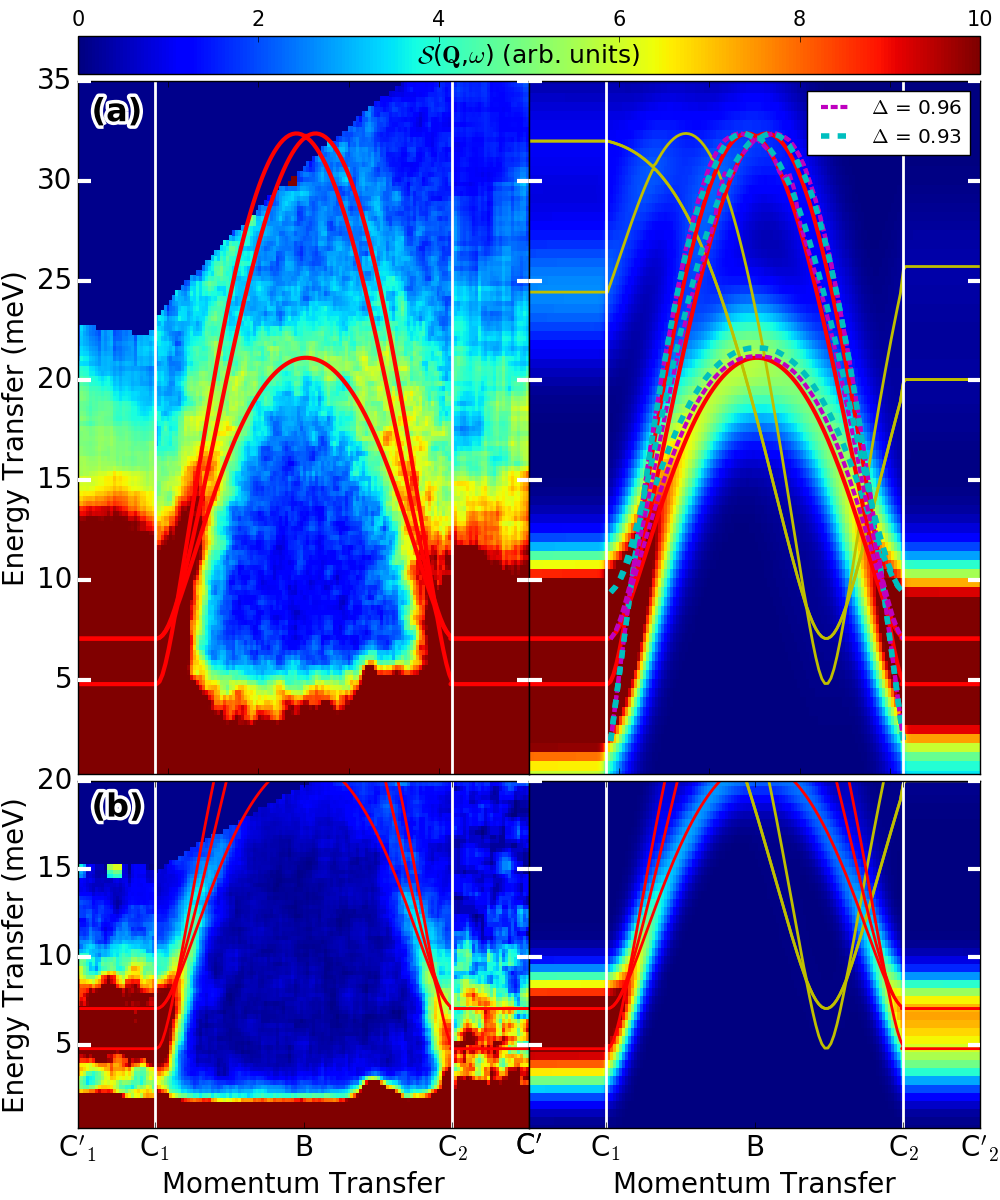}
	\caption{Comparison of measured and calculated $\cal{S}$(\textbf{Q},$\omega$) along the C$_1$BC$_2$ (C$_1$=(100), B=(1.5 0 0), C$_2$=(200)) and $c$-axis directions at C points (C$'_1$=(10$\frac{1}{4}$) and C$'_2$=(20$\frac{1}{4}$)) for incident energy settings of (a)~$E_i$~=~45~meV and (b)~$E_i$~=~24~meV. Red (yellow) lines are LSWT dispersions from the first (second) domain. Dashed colored lines represent the dispersion calculation from the XXZ model for two values of $\Delta$~=~$J_z$/$J$ as described in the main text. }
	\label{Cslices}
\end{figure}

The spin Hamiltonian for the Heisenberg model includes a nearest neighbor exchange interaction $J$ and both easy plane $K_D$ and local easy axis $K_A$ single-ion anisotropy terms were used within linear spin wave theory (LSWT) with the following equation: 
\begin{equation}
H_{LSWT} = J\sum_{<ij>}\vec{S_i}\cdot\vec{S_j}+K_D\sum_{i}(S_i^z)^2+ K_A\sum_{i}(S_i^n)^2 \tag{1} \label{LSWT}
\end{equation}
(the superscript $n$ denotes the local easy axis direction, which is simply the direction each spin is pointing in the A$_1$ magnetic structure configuration as shown in Fig.~\ref{structure}) The spin Hamiltonian shown above was diagonalized using the SpinW software package \cite{SpinW} in order to produce the spin-wave dispersions to model the experimental data. The model of the dynamical spin correlation function (or dynamical structure factor) $\cal{S}$(\textbf{Q},$\omega$) generated by SpinW is convoluted with a Gaussian function of fixed width set in accordance with the FWHM of the elastic line of the data. 

From this experiment, we have resolved the magnetic dynamics in $h$-(Lu$_{0.6}$Sc$_{0.4}$)FeO$_3$ by comparing the experimental and theoretical  $\cal{S}$(\textbf{Q},$\omega$) (Fig. \ref{Eslices}(a) and \ref{Eslices}(b)) to obtain the dispersion relations. Consequently, the exchange coupling and single-ion anisotropy terms in (Lu,Sc)FeO$_3$ were determined. This allows for the comparison of experimentally measured parameters with those recently obtained by DFT calculations \cite{das2014bulk}, which predicted that the exchange coupling constant for $h$-LuFeO$_3$ to be $J$~=~6.31~meV. We note that examination of the $E_i$~=~111~meV data shows no magnetic excitations above 40~meV. This means the full bandwith of the magnon dispersions has been captured with the $E_i$~=~45 meV data (Fig.~\ref{Eslices}(a)), allowing for an accurate experimental determination of the magnetic exchange coupling constants. 

The nearest-neighbor exchange coupling value was found to be $J$~=~4.0(2) meV from LSWT. We note that it is not possible to distinguish the effects from small amounts of trimerization (i.e. zone tripling structural distortions) with this dataset, but the average value of $J$ must be kept at 4.0(2)~meV. This means that the theoretically predicted value for the exchange constant is $\sim$58\% higher than the actual value. We note that the theoretically predicted value for the Curie-Weiss temperature ($\Theta_{CW}$~=~1525~K \cite{das2014bulk}) is similarly $\sim$53\% higher than the previously measured experimental value~\cite{PhysRevB.92.054435}. This discrepancy between experiment and theory could be from a number of sources, such as the choice of the Hubbard $U$ value in the calculations.

From the width of the split modes (separation of the red lines) at the C point (Fig. \ref{Cslices}(b)) the magnitude of the easy-plane anisotropy term was found to be $K_D$~=~0.17(1)~meV. This is almost equal \mbox{(within error)} to the equivalent theoretically predicted value, the $z$ component of a single-ion anisotropy (SIA) tensor, $\tau_{zz}$~=~0.181~meV ~\cite{das2014bulk}. From the magnitude of the spin gap (mode at $\sim$5~meV) at the C point (Fig. \ref{Cslices}(b)) the local easy axis anisotropy value was determined to be \mbox{$K_A$~=~-0.05(1)~meV.} To the best of our knowledge, the value of this local easy axis anisotropy has not been predicted from \textit{ab initio} calculations. Previous research has shown exactly how both the direction and magnitude of the SIA develops in $h$-LuFeO$_3$ \cite{SIA_orgin}.  This is similar to the origins of SIA and the Dzyaloshinskii-Moriya (DM) interaction term in BiFeO$_3$~\cite{Jeong_2014_PRL}. For this case, the theoretical explanation for what induces and modulates both the SIA and DM interactions is the trimerization distortions~\cite{das2014bulk}. 

The combined SIA and DM interactions can lead to out-of-plane spin canting, producing a weak ferromagnetic (wFM) moment along the c-axis~\cite{das2014bulk}. Since we observe no spin wave dispersion in the out-of-plane direction (see Fig.~\ref{Cslices}(b) along the C to C' directions), it may be concluded that there is no wFM component along the $c$-axis. Furthermore, from these flat magnon modes in the out-of-plane direction it can also be concluded that there is no significant interlayer exchange coupling.

It is important to point out that the A$_1$ ($\Gamma_1$) magnetic ground state configuration known for this material \cite{PhysRevB.92.054435} (which by symmetry forbids any net magnetic (wFM) moment along the c-axis) remains consistent with our data. However, this stands in contrast to previous neutron diffraction measurements on $h$-LuFeO$_3$ thin films \cite{PhysRevLett.114.217602, PhysRevLett.110.237601} and DFT results \cite{das2014bulk}, which have shown that the magnetic structure should be in the A$_2$ ($\Gamma_2$) configuration; the only one allows for a wFM component (net magnetization) of Fe$^{3+}$ spin along the c-axis. It should be noted though that the A$_1$ configuration was theoretically \cite{das2014bulk} found to be very close energetically to A$_2$. 

Based on the fact that we see no roton-like minimum at the B point in the spin-wave spectra, we can conclude that magnon-phonon coupling is either very small (at least smaller in strength than in LuMnO$_3$) or absent in this material. One possible reason for this could be the changing of relative positions between the magnon and phonon modes in this case, and therefore less magnon-phonon mode overlap required for such coupling. Another reason may be a reduction in the exchange-striction coupling, as indicated by first principles calculations of spin-lattice coupling \cite{PhysRevB.92.035107}.

We attempted to extract from the data an upper limit to any possible magnon-magnon coupling. In order to account for the possibility of anharmonic spin-waves originating from magnon-magnon interactions, we employ the Heisenberg XXZ model with $1/S$ expansions \cite{PhysRevB.79.144416} where the exchange interaction $J$ and the two-ion (easy-plane) anisotropy $\Delta$~=~$J_z$/$J$ are adjustable parameters. This is given by the following spin Hamiltonian:

\begin{equation}
H_{XXZ} = J\sum_{<ij>}\left[S_i^x S_j^x + S_i^y S_j^y + \Delta S_i^z S_j^z\right] \tag{2} \label{XXZ}
\end{equation}

The results from this spin Hamiltonian for (Lu,Sc)FeO$_3$ are summarized in Fig.~\ref{Cslices}(a) by the colored dashed lines as indicated in the legend. The parameters used for the XXZ model simulation which give the most reasonable fit are $J$~=~4.21 meV, $\Delta$~=~0.96. Any lower values of $\Delta$ (such as $\Delta$~=~0.93) cause the renormalized spectra to deviate outside of tolerances (mainly the separation width of the modes at the zone center (C point)) set by the data. 

According to Ref. \cite{PhysRevB.79.144416}, in the XXZ model the two-ion (easy-plane) anisotropy $\Delta$ directly affects the decay of coherent magnons. The parameter $\Delta$~=~0.96 found to be most consistent with the data in this case indicates that (Lu,Sc)FeO$_3$ is closer to the Heisenberg limit $\Delta \sim 1$ than in the case of LuMnO$_3$ where $\Delta$~=~0.93 \cite{oh2016}. In a 2D TLAF with a non-collinear magnetic structure, strong renormalization and decays in the Heisenberg limit \cite{PhysRevB.79.144416} are expected to be present. Therefore, the current indication seen in this dataset that (Lu,Sc)FeO$_3$ does not seem to robustly exhibit these effects is contrary to expectations. One possible reason for this may be the larger spin number $S$~=~5/2 which leads to a reduction in magnon-magnon interaction strength, consistent with these observations. This effect is clearly shown in the calculated $\cal{S}$(\textbf{Q},$\omega$) plots for $S$~=~1/2 and 3/2 cases, as reported in Mourigal \textit{et al.} \cite{PhysRevB.88.094407}, where the amount of renormalization is significantly reduced as $S$ becomes larger (though not entirely eliminated). 

\section{Conclusions}
Time-of-flight inelastic neutron scattering measurements were performed with a bulk crystal of hexagonal (Lu$_{0.6}$Sc$_{0.4}$)FeO$_3$ in order to obtain a comprehensive overview of the spin wave spectra. The recently developed SpinW \cite{SpinW} software was utilized to robustly model the magnon dispersion relations, fully taking into account the experimental effects of the sample mosaic, the two domains in the sample \cite{footnote1}, and the instrumental resolution function. We have determined the values for the relevant magnetic interactions in this system, which allows for an accurate comparison between experimental results and theoretical calculations. For the nearest neighbor exchange coupling $J$, we find that the experimental value is $\sim$~2/3 in comparison to predictions by DFT calculations \cite{das2014bulk}. We have also confirmed the ground state magnetic structure is of A$_1$ ($\Gamma_1$) type. With the previously reported exchange parameters for LuMnO$_3$ \cite{oh2016}, it is now possible to compare exactly how the change of the transition metal cation from Mn$^{3+}$ to Fe$^{3+}$ adjusts the magnetic properties of these ABO$_3$ compounds.  This study shows that the overall energy scale of the spin wave spectra is twice as large for LuFeO$_3$ ($S$~=~5/2) as it is for LuMnO$_3$ ($S$~=~2) and that magnon-magnon/phonon couplings are reduced in LuFeO$_3$. 

\section{Acknowledgments}
We wish to thank Kyungsoo Kim for helpful discussions. Work at the IBS CCES (South Korea) was supported by the research program of the Institute for Basic Science (IBS-R009-G1). Work at Rutgers University was supported by the DOE under Grant No. DOE: DE-FG02-07ER46382. We acknowledge the Rutherford Appleton Laboratory for access to the ISIS Neutron Source.

\bibliography{LuFeO3_ref}

\newpage
\onecolumngrid
\section{Supplemental Information}

Fig~\ref{Qcuts} below shows the constant-\textbf{Q} cuts through the data at points along the high symmetry directions indicated in Fig~\ref{structure}(c) and Fig~\ref{Eslices}. From these plots the peak centers and peak widths may be more readily discerned. The peak fits shown by the black circles are  plotted commensurately at their precise \textbf{Q} positions in Fig~\ref{Eslices}. 
\begin{figure}[!h]
	\centering
	\includegraphics[width=1.0\columnwidth,clip]{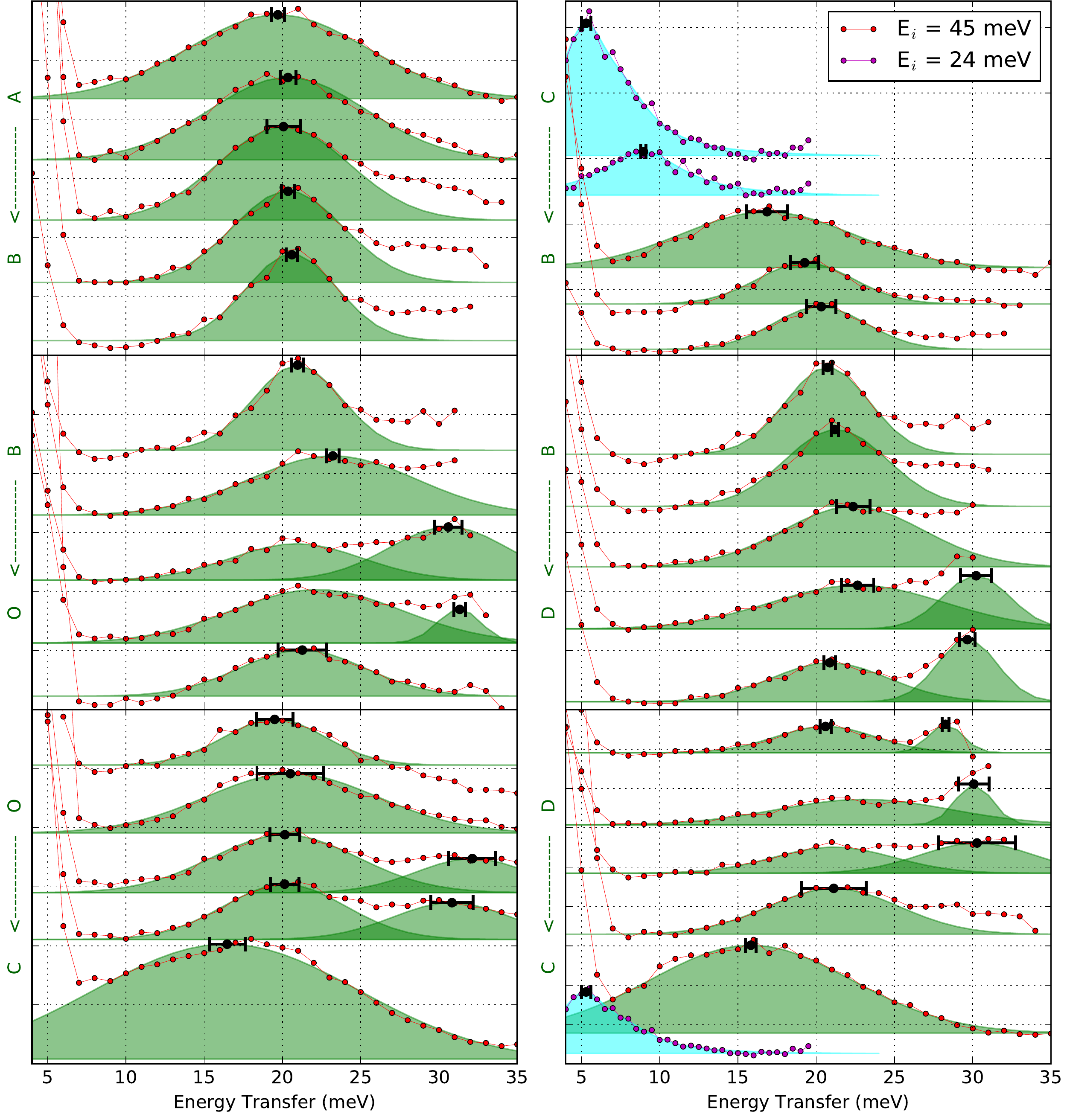}
	\caption{ Constant-\textbf{Q} cuts through the INS data at the \textbf{Q} positions shown by the corresponding black circles in Fig~\ref{Eslices}. Red (purple) lines denote  cuts taken from the E$_i$=~45~meV (E$_i$=~24~meV) datasets. The black circles and error bars indicate the positions of fitted peak centers and the uncertainty of peak center fits. Shaded regions are the fitted peak areas. }
	\label{Qcuts}
\end{figure} 

\end{document}